\documentclass[letterpaper, 10 pt, conference]{ieeeconf}  

\IEEEoverridecommandlockouts                              
\usepackage[all]{xy}
\usepackage{color}
\usepackage{subfigure}
\usepackage{bm}
\usepackage{graphicx}
\usepackage{color}
\usepackage{wrapfig}
\usepackage{epsf}
\usepackage{times,mathptm}
\usepackage{url}
\usepackage{amssymb, amsmath}

\title{\LARGE\bf
Exact and Efficient Algorithm to Discover Extreme Stochastic Events\\
in Wind Generation over Transmission Power Grids}

\author{Michael Chertkov, Mikhail Stepanov, Feng Pan, and Ross Baldick
\thanks{
The work of MC and FP at LANL was carried out under the auspices of the
National Nuclear Security Administration of the U.S. Department of
Energy at Los Alamos National Laboratory under Contract No.
DE-AC52-06NA25396. The work of MC, MS and FP was funded in part by DTRA/DOD under the grant BRCALL06-Per3-D-2-0022 on ``Network Adaptability from WMD Disruption and Cascading Failures".
The work of MS is also partially supported by NSF grant DMS-0807592. RB is funded, in part, by the Department of Energy under Contract 09EE0001385.}
\thanks{M. Chertkov is with Theory Division \& Center for Nonlinear Studies at LANL,
Los Alamos, NM 87545 and also with New Mexico Consortium, Los Alamos, NM 87544 {\tt\small chertkov@lanl.gov}}
\thanks{M. Stepanov is with Department of Mathematics, University of Arizona,
Tucson, AZ 85721 {\tt\small stepanov@math.arizona.edu}}
\thanks{F. Pan is with Decision Division at LANL, Los Alamos, NM 87545 {\tt\small fpan@lanl.gov}}
\thanks{R. Baldick is with Department of Electrical and Computer Engineering
Engineering, The University of Texas at Austin, Austin, TX 78712 {\tt\small baldick@ece.utexas.edu}}
}

\begin{document}

\maketitle
\thispagestyle{empty}
\pagestyle{empty}

\begin{abstract}
In this manuscript we continue the thread of [M. Chertkov, F. Pan, M. Stepanov, Predicting Failures in Power Grids: The Case of Static Overloads, IEEE Smart Grid 2011] and suggest a new algorithm discovering most probable extreme stochastic events in static power grids associated with intermittent generation of wind turbines. The algorithm becomes EXACT and EFFICIENT (polynomial) in the case of the proportional (or other low parametric) control of standard generation, and log-concave probability distribution of the renewable generation,  assumed known from the wind forecast. We illustrate the algorithm's ability to discover problematic extreme events on the example of the IEEE RTS-96 model of transmission with additions of $10\%, 20\%$ and  $30\%$ of renewable generation. We observe that the probability of failure may grow but it may also decrease with increase in renewable penetration, if the latter is sufficiently diversified and distributed.
\end{abstract}

\section{Introduction}

In progress to  becoming smarter, the power grid of today is undergoing multiple transformations \cite{09EDR,09EERE}. One of the envisioned changes consists in replacing a significant portion of the traditional fossil thermal plants by renewable generation \cite{20wind}, in particular by solar and wind farms. This task,  motivated by ecological and political reasons, will not be a simple substitution. The renewable sources of energy, will also be much less predictable and thus much more difficult to control. Both wind and solar fluctuate temporally and spatially. Even when forecasted,  the renewable generation can be described only in probabilistic terms,  suggesting that the existing toolbox of power engineering, which is largely deterministic, needs to be upgraded with computationally more challenging probabilistic tools. For any configuration of aggregated loads, which stays roughly constant for tens of seconds or even minutes, one ought to consider an ensemble of possible configurations of renewable generation undergoing significant and unpredictable changes during the same time. This uncertain (but probabilistically predictable) configuration of renewable generation will be complemented by standard adjustable generation. The adjustment is required to stabilize the system and complete matching between requested load and total generation. Many configurations from this ensemble of matchings will be feasible, i.e. not violating transmission constraints. However,  there will always be some number of probabilistically rare but strong fluctuations of the renewable sources for which an instantaneous adjustment of generation to loads will be problematic,  as resulting in overload of some number of lines. Discovering extreme stochastic events of this rare but damaging type becomes an important practical problem which requires fast and largely non-existent algorithmic solutions.

This manuscript is devoted to resolving (at least some part of) the aforementioned challenge. We assume that configuration of loads over the network is fixed,  while generation is split into two parts: (a) the bigger, consisting of conventional generation, controlled and adjustable as a group, for example via common re-scaling
(roughly representing, for example, so-called ``droop'' speed control with the proportion of each generator response to the frequency variation fixed, or representing ``regulation'' response when deployed regulation is also in proportion to the frequency variation); and (b) renewable generation, say coming from wind farms, uncertain and not controlled,  described in terms of a probability distribution function,  assumed known from a forecast. We are interested to discover the most probable configuration(s) of the wind (more generally any fluctuating renewable source) which is troublesome,  in terms of possibly violating transmission and/or controlled generation constraints, when no additional control efforts, such as curtailment of wind generation,  load shedding, line switching, etc, are in place. With diversity of wind production across multiple wind sites,
the extreme stochastic events will not be frequent, thus making the problem of finding the rare configuration causing the trouble as challenging as finding a needle in a haystack. To discover the extreme stochastic events we exploit and develop further the approach, originally introduced in theoretical physics, see e.g. \cite{64Lif,66ZL,96FKLM,97Che} then used to analyze performance of error-correction codes \cite{08CS}, and recently applied to predict static failures in power grids associated with fluctuations in loads \cite{11CPS}.  We call this rare but most probable troubled instance of the wind the instanton. In a significant technical improvement in comparison with \cite{11CPS} this manuscript suggests  a direct way, which is Exact and Efficient (polynomial in the size of the grid),  to explore the structure of the problem for finding the instanton(s). This algorithmic improvement is achieved via mapping the instanton problem to minimization of a convex function, characterizing forecasted distribution of the wind, over the exterior of a tractable polytope, describing feasible solutions of the power flow equations.

The material in the manuscript is organized as follows. We give some technical background,
formulate the problem and briefly describe history of the instanton methodology in Section \ref{sec:Formulation}. Section \ref{sec:Exact} formulates the main theoretical result of the manuscript: exact and efficient algorithm for finding most probable wind extreme stochastic event for given configuration of the grid. The performance of the algorithm is illustrated in Section \ref{sec:Numerical} on example of the IEEE RTS96 system with $10\%, 20\%$ and $30\%$ of renewable penetration. The results are summarized and discussed in Section \ref{sec:Summary},  where we also discuss future research challenges.

\section{Quasi-Static Power Flows, Wind Modeling and Formulating the Instanton Problem}
\label{sec:Formulation}

Modern power grid is always in motion. Moreover many of the changes which take place on multiple time scales are inherently uncertain, being probabilistic in nature.  With sufficient penetration of the renewable sources, wind turbines bring a particularly important source of fluctuations into the modern power grid. Production of wind can be forecasted only to a degree.  Even with a perfect forecast,  and due to a turbulent nature of the wind, one ought to describe output of the wind turbine integrated directly to the power grid (without curtailing) in probabilistic, rather than deterministic, terms. Relevant times scales,  where fluctuations of wind generation dominate other sources of uncertainty (in particular these associated with fluctuations of demands), are in the range from tens of seconds to tens of minutes. At these temporal scales transient (sub-second) phenomena are already settled and quasi-static description of the power flows is appropriate \cite{94Kun,96WW}. Taking a standard Direct Current (DC) power flow approximation (well justified for transmission network,  where resistivity of power lines is significantly smaller that respective inductance,  voltage variations are small and change in phase between neighboring busses is small too) one arrives at the following relations between the vector of injected/consumed real powers, ${\pmb P}$, and the vector of phases, ${\pmb\varphi}$, over the power grid defined as a graph,  ${\cal G}=({\cal G}_0,{\cal G}_1)$ (${\cal G}_0$ and ${\cal G}_1$ represent the set of nodes and the set of edges respectively),
written in the form of conditions
\begin{eqnarray}
&& COND_{flow}=\Biggl({\pmb B}{\pmb\varphi}={\pmb P},\ \&\ \varphi_0=0\Biggr),\label{flow_cond}\\
&& {\pmb B}=\left\{\begin{array}{cc} 0,& \{i,j\}\notin {\cal G}_1\\ -1/x_{ij},& \{i,j\}\in{\cal G}_1\\
\sum_{k}^{\{i,k\}\in{\cal G}_1}x_{ik}^{-1},& i=j\end{array}\right.,\label{B}\\
&&{\pmb P}=\left(\begin{array}{cc} -d_i,& i\in{\cal G}_d\\ \rho_i,& i\in{\cal G}_r\\
\alpha p_i,& i\in{\cal G}_g\\
0,& i\notin {\cal G}_d\cup {\cal G}_r\cup {\cal G}_g\end{array}\right).
\label{P}
\end{eqnarray}
In the remainder of the paragraph we detail notations, and underlying notions, assumed in Eq.~(\ref{P}).
${\pmb B}$ in Eq.~(\ref{flow_cond}) is the matrix of the graph Laplacian constructed from line inductive reactances,  ${\pmb x}$,  and the second condition in Eq.~(\ref{flow_cond}) sets the phase at an arbitrarily chosen (zero) node to zero.  Different components of the vector, ${\pmb P}$, associated with nodes of the graph, are  split in Eq.~(\ref{P}) into four groups: set of demand/consumption nodes, ${\cal G}_d$; set of renewable nodes, ${\cal G}_r$; set of standard generation nodes, ${\cal G}_g$;  and the remaining set of other (junction) nodes. Any component of the demand is assumed negative and fixed (on the time scale of interest measured in seconds-to-minutes). The renewable generation fluctuates according to the forecasted probability distribution
\begin{eqnarray}
{\cal P}({\pmb \rho})\sim \exp\left(-{\cal S}({\pmb \rho})\right),
\label{P-rho}
\end{eqnarray}
where  ${\cal S}({\pmb \rho})$ is a known convex function of its multi-dimensional argument, achieving minimum at the most probable, equilibrium, configuration of the wind generation, ${\pmb \rho}_0$. This distribution function represents statistics of the wind fluctuations, collected over the time interval when loads do not change or change very little. (The assumption of ${\cal S}$ convexity is reasonable for a sufficiently wide, and thus most probable, vicinity of the maximum output, in particular for popular modeling of the wind statistics via the Weibull function \cite{Weibull}.)
Standard (controllable) generation adjusts to match the imbalance between total consumption and total renewable generation (assuming that the loads stay constant). The adjustment is relatively fast (instantaneous within the quasi-static description) and we will simplify by assuming that the mismatch is shared between the controllable generators in some pre-defined, automatic fashion. In particular, we will consider the case where the adjustment for each generator is proportional to its nominal generation, so that $\alpha$ entering Eq.~(\ref{P}), is
\begin{eqnarray}
\alpha=\frac{\sum_{i\in {\cal G}_d} d_i-\sum_{i\in {\cal G}_r}\rho_i}{\sum_{i\in{\cal G}_g} p_i},
\label{alpha}
\end{eqnarray}
and it is the only degree of freedom on the standard generation side which reacts absorbing changes in the renewable generation, ${\pmb \rho}=(\rho_i|i\in{\cal G}_r)$. We assume that $\alpha$ is set to unity at the equilibrium configuration, ${\pmb \rho}_0$. Then,  $(p_i|i\in {\cal G}_g)$ constitutes output of the controllable generation preset in the beginning of the time interval of interest.  Normally,  this redispatch of the controllable generation is the output of the Optimum Power Flow (OPF) analysis, accounting for diversity in the generation cost at different sites, and executed periodically on the scale ranging from minutes to tens of minutes. The model of Eq.~(\ref{alpha}) schematically represents either droop control or regulation response, although details
of both of these control modes differ from our model in details.

One expects that a feasible Power Flow (PF) solution, representing a perturbed OPF, satisfies, in addition to the basic power flow relations (\ref{flow_cond}), the following set of transmission (thermal) conditions representing line constraints on the amount of power which can flow safely trough the lines (without overheating or damage)
\begin{eqnarray}
&& COND_{edge}=\Biggl(\forall \{i,j\}\in{\cal G}_1:\quad |\varphi_i-\varphi_j|\leq x_{ij}u_{ij}\Biggr),
 \label{edge_cap_cond}
\end{eqnarray}
where $u_{ij}$ is the line $\{i,j\}$ rating.
One also assumes that all the generators included in the proportional control are within their capacity bounds,  which translates into the following cumulative constraints on the proportional control coefficient
\begin{eqnarray}
 && COND_{power}=\Biggl(\underline{\alpha}\leq \alpha\leq \overline{\alpha}\Biggr),\label{power_cap_cond}
\end{eqnarray}
where $\underline{\alpha}$ and $\overline{\alpha}$ are defined in accordance with the maximum low and minimum high constraints,  respectively, over individual controllable generators. In fact, typical control modes would allow for some generators reaching their limits; however, we simplify that issue here.

Even though the most probable configuration of the wind generation corresponds to the power flow solution of $COND_{flow}$, which is safely within the feasibility region of $COND_{edge}\cup COND_{power}$, other less probable configurations of wind can violate one or more of the feasibility constraints. Naturally,  we are first of all interested to discover the most probable instance of the wind, the instanton, which lies outside of the feasibility region:
\begin{eqnarray}
&& {\pmb\rho}_{inst}=\left.\mbox{argmin}_{{\pmb \rho}}{\cal S}({\pmb \rho})
\right|_{{\pmb \rho}\notin {\cal D}_{int}},\label{inst}\\
&& {\cal D}_{int}\equiv \mbox{Projection}\left(COND({\pmb\varphi},{\pmb \rho},\alpha)\right)_{\pmb \rho},
\label{D_int}\\
&& COND\equiv COND_{flow}\cup COND_{edge}\cup COND_{power},
\label{COND}
\end{eqnarray}
where $\mbox{Projection}\left(COND({\pmb\varphi},{\pmb \rho},\alpha)\right)_{\pmb \rho}$ is the projection
of the polytope $COND$ to the ${\pmb\rho}$-space. Therefore,  by construction ${\cal D}_{int}$ is also a polytope. Alternatively we can rewrite Eq.~(\ref{inst}) as
\begin{eqnarray}
&& {\pmb\rho}_{inst}=\left.\mbox{argmin}_{{\pmb \rho}}{\cal S}({\pmb \rho})
\right|_{{\pmb \rho}\in {\cal D}_{ext}},
\label{var1}
\end{eqnarray}
where ${\cal D}_{ext}$ is a non-convex set,  defined as an exterior of the convex set (polytope) ${\cal D}_{int}$, i.e. ${\cal D}_{ext}=R_+^{|{\cal G}_u|}\setminus {\cal D}_{int}$. Eq.~(\ref{var1}) states succinctly the instanton problem addressed in this manuscript.

In more general formulations, not yet utilizing special structure of the optimization domain ${\cal D}_{ext}$ specific to our problem, the instanton problem (\ref{var1}) can be solved within the machinery of non-convex optimization methods.  For example, and as described in details in \cite{04CCSV,05SCCV,11CPS}, one may search for the minimum of ${\cal S}({\pmb\rho})$ with the help of a general-purpose optimization technique, specifically the downhill simplex (or ``amoeba'') method \cite{AMOEBA,NumRec}. Since the optimization domain  is not concave, we expect to find many (candidate) instanton solutions. Each initialization of the instanton-amoeba could lead to a new instanton. The initialization selects a simplex, built on $N_d+1$ feasible points (where $N_d$ is the dimensionality of the ${\pmb\rho}$ space, which is equal in our case to the number of renewable generators) from the error-surface separating ${\cal D}_{int}$ and ${\cal D}_{ext}$. Then the instanton-amoeba method evolves the simplex, via a sequential set of shifts, contractions and extensions, towards its eventual collapse to a local minimum of ${\cal S}({\pmb \rho})$. Different random initiations will sample the space of instantons, thus generating the so-called instanton spectrum describing the frequency of a given instanton occurrence and also suggesting an estimation for the ordered list (with respect to their probability of occurrence and frequency) of top instantons.  Repeated infinite number of times, the sampling would output the most probable instanton. However given that the number of initiations will be finite in reality, the most probable one (of the finite number of instantons found) gives a heuristic estimate from below for the probability of the most probable instanton. To ensure sampling quality one needs to continue random initiations till the most probable instantons would appear multiple number of times. (Typically, this require hundreds of initiations for the network measured in hundreds of nodes, of the type discussed below in Section \ref{sec:Numerical}.)

However, the specific structure of our problem (\ref{var1}) allows a much faster and moreover exact resolution,  than the one provided by the general purpose but computationally heavy instanton-amoeba method of \cite{04CCSV,05SCCV,11CPS}. As shown in the next Section, very specific features of ${\cal D}_{ext}$,  associated with the linear nature of the DC power flow and also with the single-parametric and linear control of fluctuations on the standard generation side, allow to solve Eq.~(\ref{var1}) efficiently. The new approach, describe below, offers a significant algorithmic improvement in comparison with the general approach of the instanton-amoeba type.

\section{Exact and Efficient Algorithm to Discover the Instanton(s)}
\label{sec:Exact}

Given Eqs.~(\ref{flow_cond}) and Eq.~(\ref{alpha}), one can express ${\pmb \varphi}$ and $\alpha$ via $\rho$,  thus arriving at the following explicit (tractable polytope) expression for  ${\cal D}_{int}$:
\begin{eqnarray}
&& \hspace{-0.5cm}{\cal D}_{int}=\overline{COND}_{edge}\cup \overline{COND}_{power},
\label{D-int}\\
&& \hspace{-0.5cm}\overline{COND}_{edge} \label{tilde_edge_cap_cond}\\
&& \hspace{-0.5cm}=\Biggl(\forall \{i,j\}\in{\cal G}_1:\ |(\tilde{\pmb B}{\pmb P})_i-(\tilde{\pmb B}{\pmb P})_j|\leq x_{ij}u_{ij}\Biggr), \nonumber\\
&& \hspace{-0.5cm}\overline{COND}_{power}\label{tilder_power_cap_cond}\\
&& \hspace{-0.5cm}=\Biggl(\!\sum_{i\in {\cal G}_d} d_i\!-\!\underline{\alpha}\sum_{i\in{\cal G}_c} p_i   \geq \sum_{i\in {\cal G}_u}\rho_i\geq \sum_{i\in {\cal G}_d} d_i\!-\!\overline{\alpha}\sum_{i\in{\cal G}_c} p_i\!\Biggr),\nonumber
\end{eqnarray}
where matrix  $\tilde{\pmb B}$ is the quasi-inverse of ${\pmb B}$ accounting for the $\varphi_0=0$ constraint and thus regularizing the only zero eigenvalue of ${\pmb B}$.

With this explicit formulation of ${\cal D}_{int}$ the instanton problem (\ref{var1}) becomes tractable, as it reduces to a minimum over a tractable set of convex problems:
\begin{eqnarray}
&& \min_{a=1,\cdots,K}\{{\cal M}_a\},\quad
{\cal M}_a=\left.\min_{{\pmb\rho}}{\cal S}({\pmb \rho})\right|_{\underline{c}_a \cup ({\cal D}_{int}\setminus c_a)},\label{M}
\end{eqnarray}
where $K=|\overline{COND}_{edge}\cup \overline{COND}_{power}|$; $c_a$ stands for any of the inequality constraints in $\underline{COND}_{edge}\cup \underline{COND}_{power}$;
and $\underline{c}_a$ is the saturated version of $c_a$ by replacing inequality with equality. When the feasibility set of a sub-problem $a$ in Eq.~(\ref{M}) is empty we formally set ${\cal M}_a$ to infinity. Eq.~(\ref{M}) is computationally tractable because it splits into $K$ convex optimizations. The proof of the transition from Eq.~(\ref{var1}) to Eq.~(\ref{M}) is straightforward and it simply relays on testing the faces of ${\cal D}_{int}$ sequentially. (See, e.g., detailed discussion of similar problem in \cite{97Bri}.)
For any of the internal minimizations in Eq.~(\ref{M}), which is feasible,  having one facet of ${\cal D}_{int}$ saturated is guaranteed (by construction) and unless there is a degeneracy in ${\cal S}({\pmb \rho})$ there will be only one saturated facet per any minimization problem ${\cal M}_a$.

Two comments are in order.  First,  our scheme is general, and can account for other types of the generation control besides adjusting generation linearly in response to the variation in ${\pmb\rho}$. However, one should also expect that the difficulty of the generalized problem will grow exponentially with the number of the control degrees of freedom. This exponential explosion is associated with the fact that projection of the respective generalization of $COND({\pmb\varphi},{\pmb \rho},{\pmb\alpha})$, where ${\pmb\alpha}$ is now multi-parametric, will generate generalized ${\cal D}_{int}$, which is much more complex as described via an exponentially large (in the dimensionality of ${\pmb \alpha}$) number of contraints/inequalities. This is in spite of the fact that the original polytope, $COND({\pmb\varphi},{\pmb \rho},{\pmb\alpha})$,  is tractable. (Indeed, it is well known that projection of a tractable polytope along a subspace results in a polytope characterized,  in the worst case, in terms of the set of constraints which is exponentially large in the size of the subspace. See \cite{96FP,00Avi} for examples and related algorithmic discussions.) Second, given that the instanton problem is split into $K$ tractable optimizations in Eq.~(\ref{M}),  one also finds not only the instanton itself (absolute minimum) but also the ranked list of other extreme stochastic events associated with other faces of ${\cal D}_{int}$.  We will call these other instantons, second-, third- etc according to their ranking in the derived hierarchy. This ranking is useful to discover the list of extreme stochastic events.
Note,  however, that strictly speaking this ranking does not necessarily corresponds to the actual ranked list of all possible extreme stochastic events. There are two  reasons for that.  First,  any point from a small continuous vicinity of the (top ranked) instanton, sitting at the same face of ${\cal D}_{int}$ as the instanton, will have a lower ${\cal S}({\pmb\rho})$ weight than other instantons. (This is under assumption that the situation is not degenerate and the instanton corresponds to an interior point of the face.) Second, one may find a (discrete) configuration, which will be second in ranking within a minimization ranked $k_1$ in Eq.~(\ref{M}), but will still have a lower ${\cal S}({\pmb\rho})$ weight than the top result of another minimization ranked $k_2$, even when $k_1<k_2$.

\section{Numerical Example}
\label{sec:Numerical}

\begin{figure}
 \centering
\includegraphics[width=2.5in,page=1]{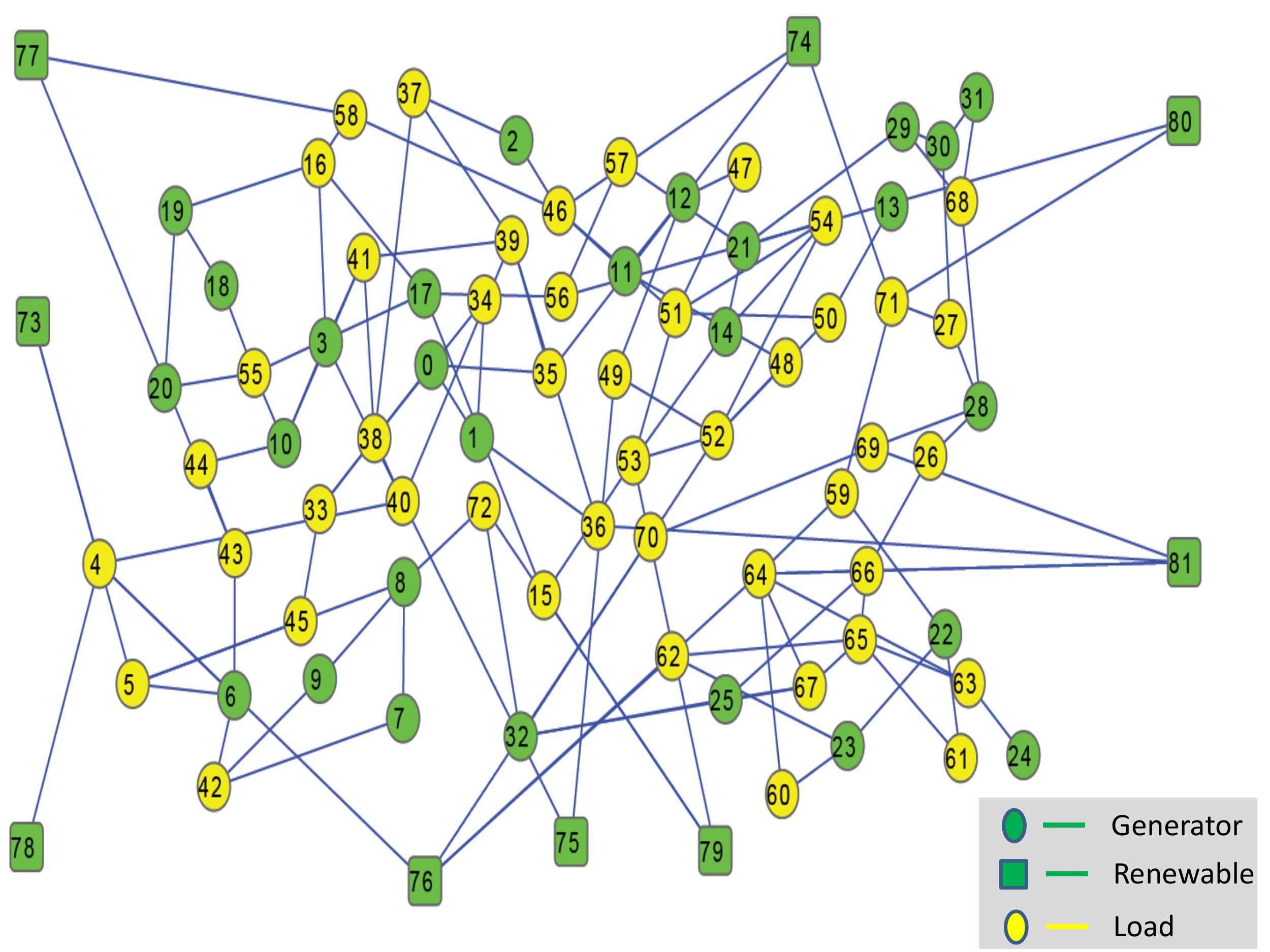}
\caption{Graph of the IEEE RTS-96 model with nine new renewable generators (\#73-\#81) added.}
\label{fig:graph}
\end{figure}

\begin{figure}
 \centering
 \subfigure[
 Configuration of renewable generation for the top three instantons.
 ]{\includegraphics[width=2.5in,page=2]{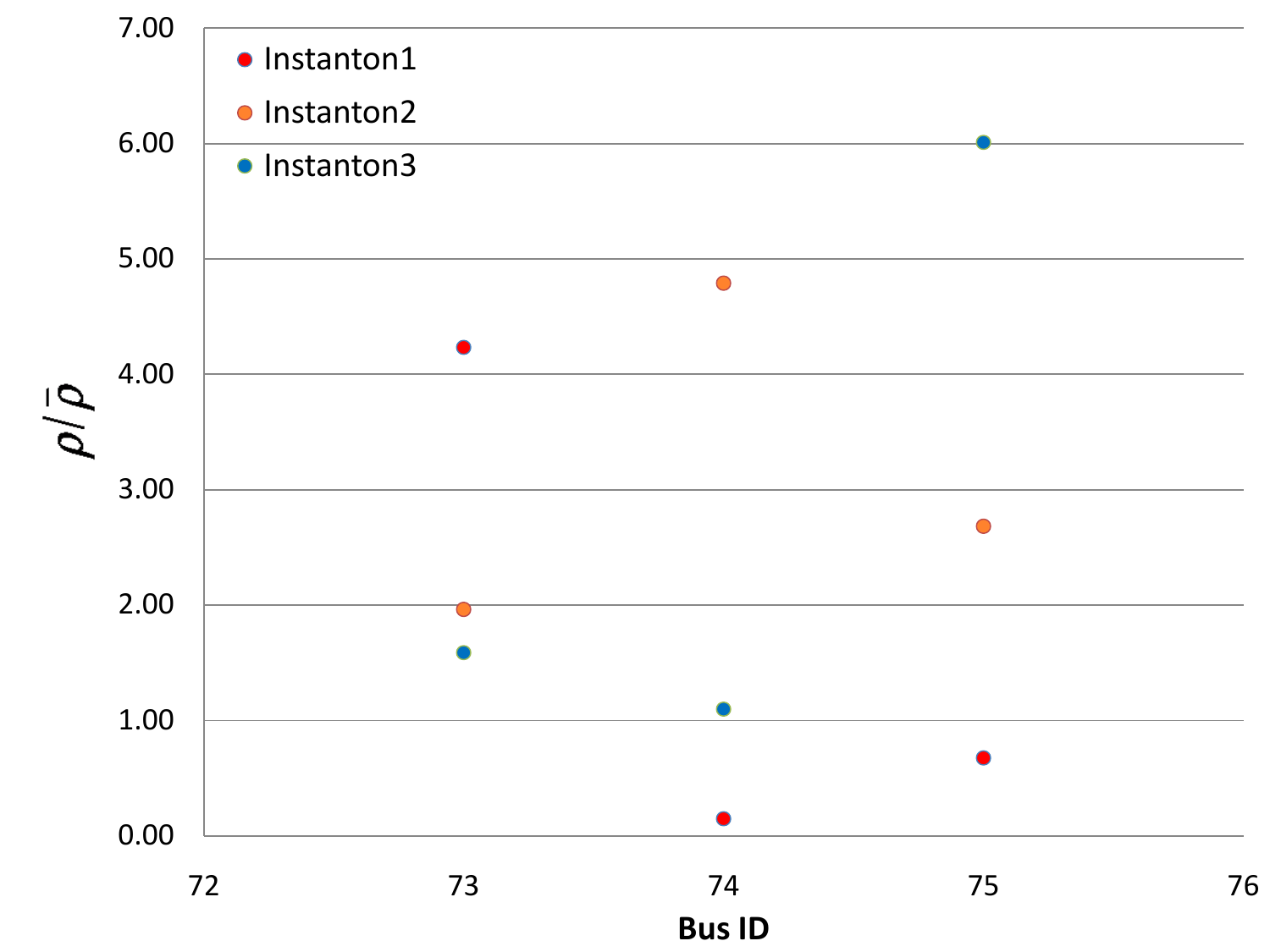}\label{fig:3a}}
 \subfigure[
 Top instantons are shown. Vertexes marked in red/orange/blue colors are of the most stressed renewable generator from the instanton output.  Edges marked red/orange/blue colors are the saturated ones for the respective instantons.
 ]{\includegraphics[width=2.5in,page=3]{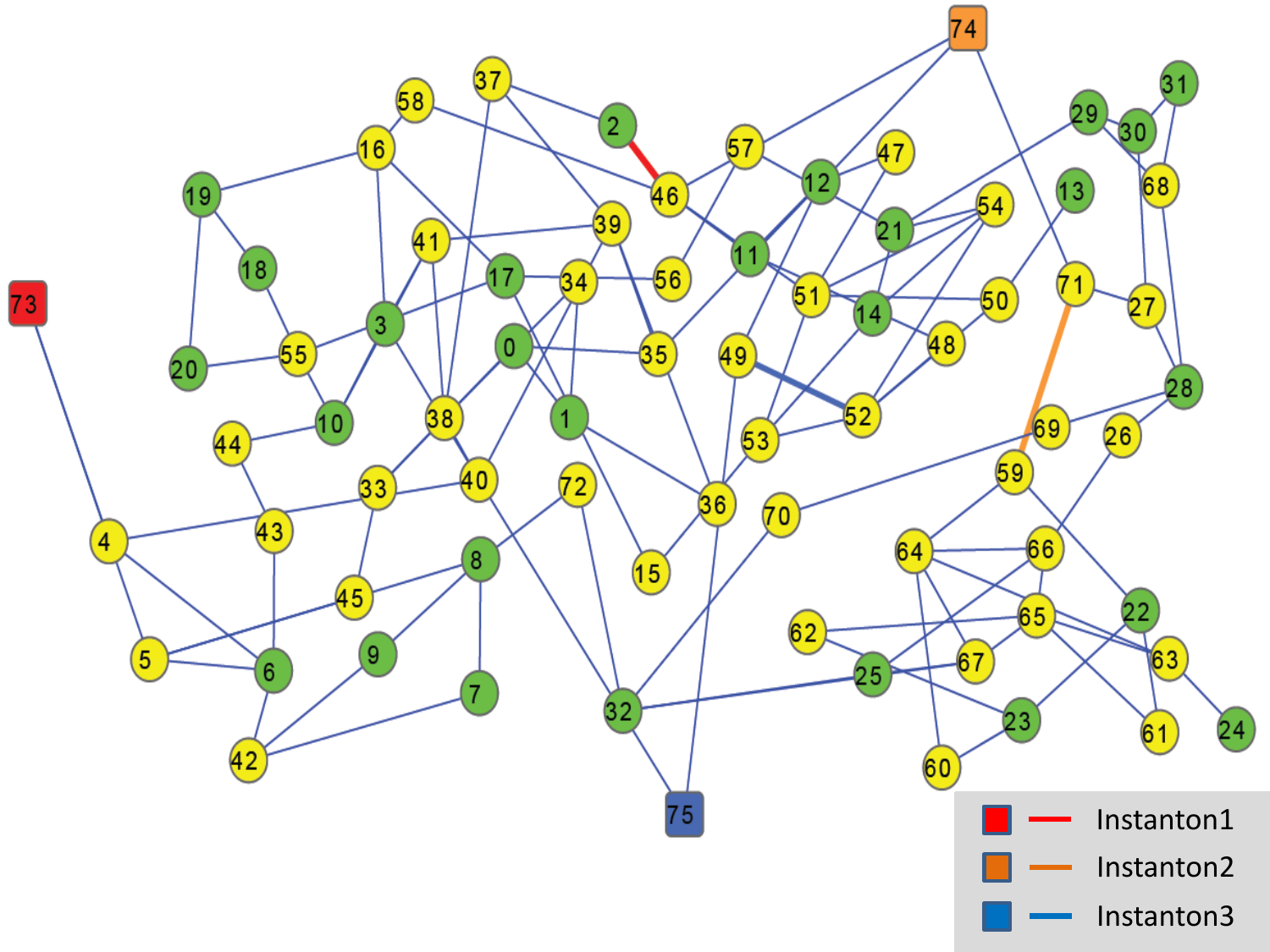}\label{fig:3b}}
 \caption{Instanton(s) for base configuration + $3$ renewable generators: $10\%$ of renewable penetration in average production. The instantons are ordered according to their cost values, $S({\pmb\rho})$.}
\label{fig:3}
\end{figure}

\begin{figure}
 \centering
 \subfigure[
 Configuration of renewable generation for the top three instantons.
 ]{\includegraphics[width=2.5in,page=4]{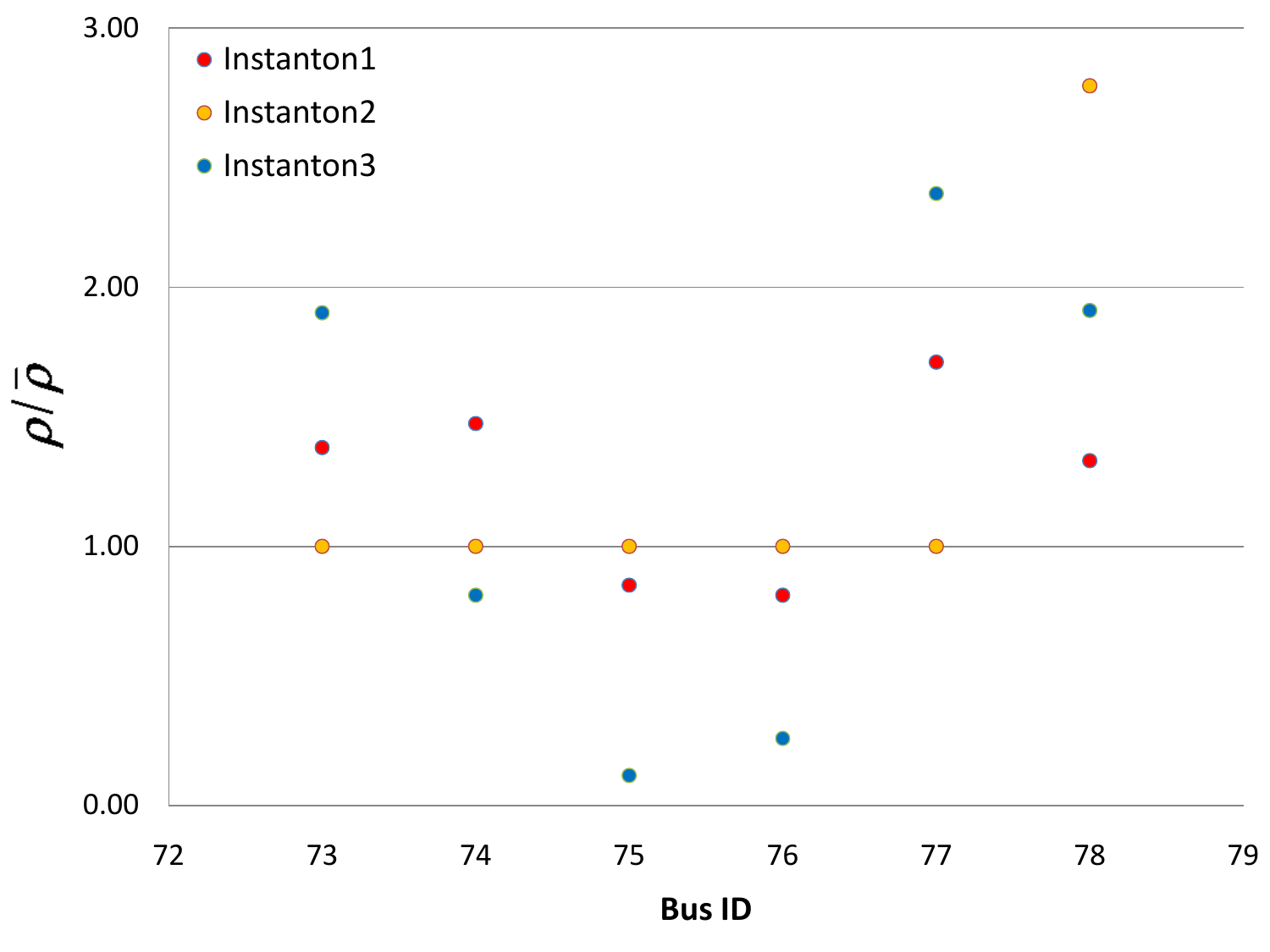}\label{fig:6a}}
 \subfigure[
 Top instantons are shown. Vertexes marked in red/orange/blue colors are of the most stressed renewable generator from the instanton output. (Bus $\#77$ is the most stressed in the instanton \#1 and the instanton \#3.) Edges marked red/orange/blue colors are the saturated ones in the respective instantons.
 ]{\includegraphics[width=2.5in,page=5]{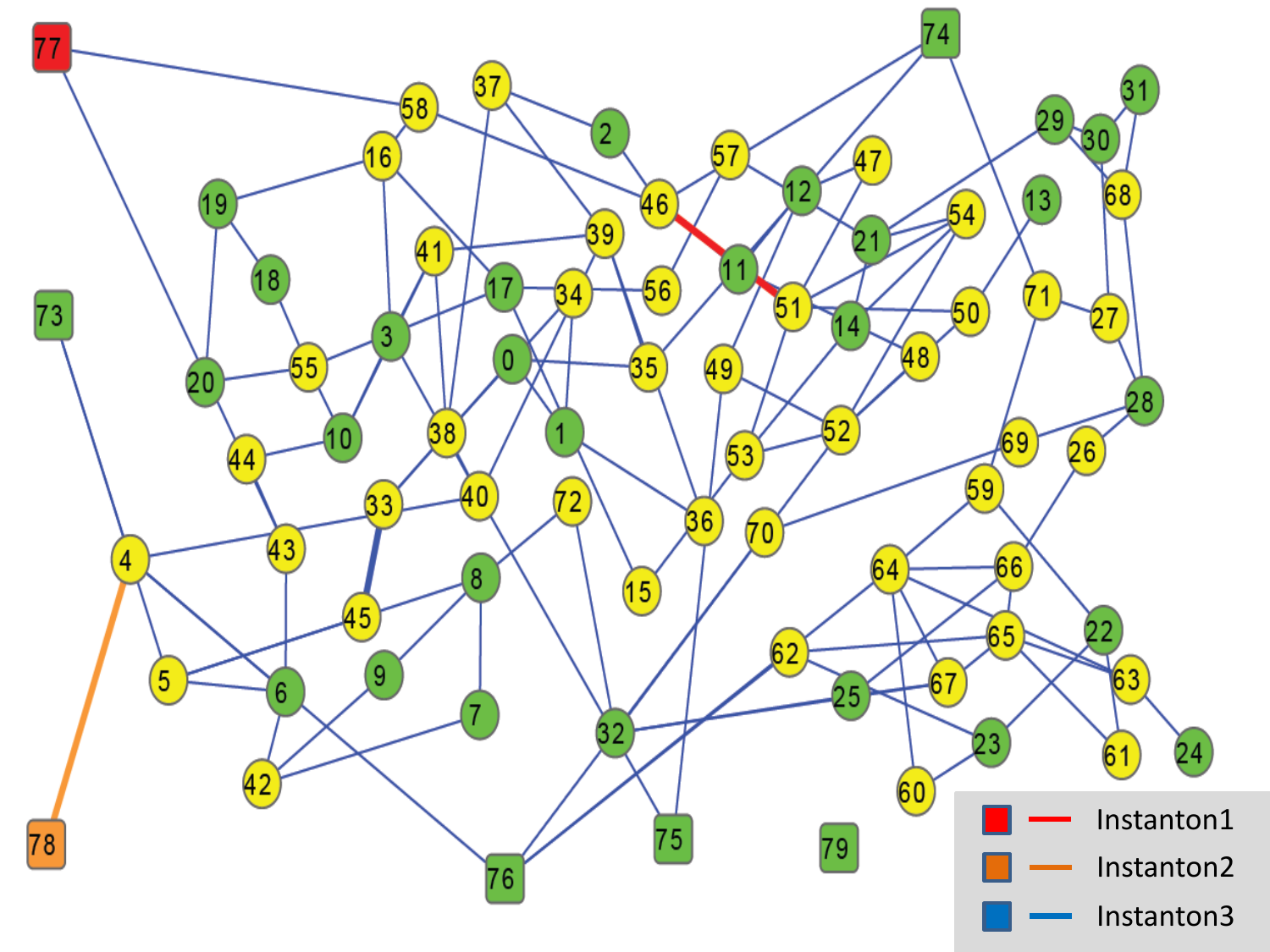}\label{fig:6b}}
 \caption{Instanton(s) for base configuration + $6$ renewable generators: $20\%$ of renewable penetration in average production. The instantons are ordered according to their cost values, $S({\pmb\rho})$.}
\label{fig:6}
\end{figure}

\begin{figure}
 \centering
 \subfigure[
 Configuration of renewable generation for the top three instantons.
 ]{\includegraphics[width=2.5in,page=6]{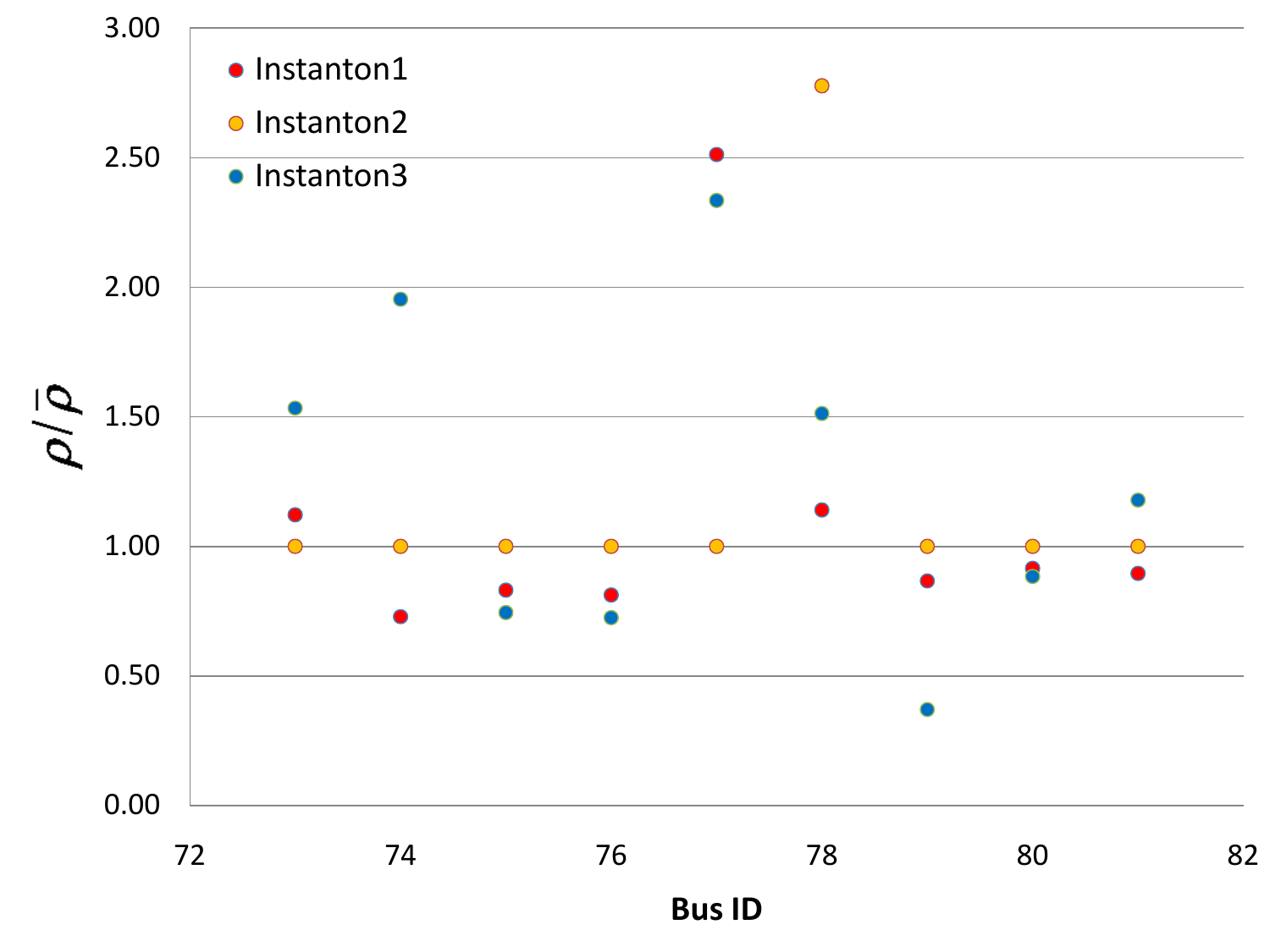}\label{fig:9a}}
 \subfigure[
 Top instantons are shown. Vertexes marked in red/orange/blue colors are of the most stressed renewable generator from the instanton output. (Bus $\#77$ is the most stressed in the instanton \#1 and instanton \#3.) Edges marked red/orange/blue colors are the saturated ones in the respective instantons.
 ]{\includegraphics[width=2.5in,page=7]{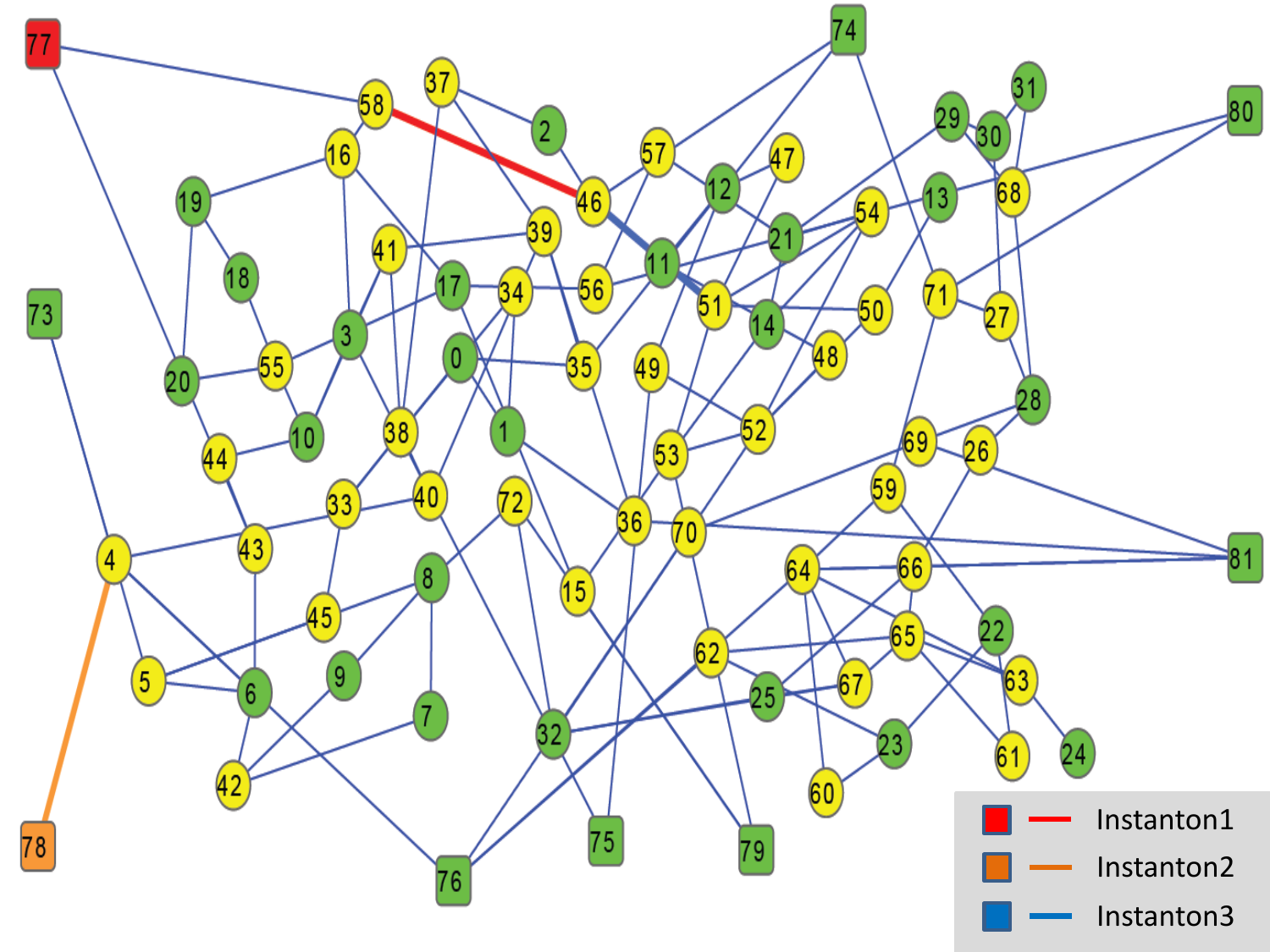}\label{fig:9b}}
 \caption{Instanton(s) for base configuration + $9$ renewable generators: $30\%$ of renewable penetration in average production. The instantons are ordered according to their cost values, $S({\pmb\rho})$.}
\label{fig:9}
\end{figure}

We test the algorithm on the standard IEEE RTS-96 model \cite{96-RTS} extended with renewables.  The results are shown in Figs. \ref{fig:graph},\ref{fig:3},\ref{fig:6},\ref{fig:9}) and commented upon below. To imitate effect of renewables, we took the base configuration of the model (standard generators, loads and transformer nodes are kept according to the data from \cite{96-RTS}).  Then, we add $3,6$ and $9$ additional nodes, for new wind generators  and connect each of them to some number of other randomly selected nodes, where number of connections per new node is distributed according to the degree distribution of generators in the base configuration. To facilitate comparison, we also create the three configurations sequentially,  such that the bigger one is built on the top of the smaller one. The new graph is shown in Fig.~(\ref{fig:graph}). The capacities of added lines are chosen equal (and roughly correspond to the median capacity of the existing lines).  Inductances of the new lines are distributed uniformly in a median range of inductances of the base case. We choose $\underline{\alpha}=0$ and $\overline{\alpha}=2$. We pick the simplest possible model for statistics of renewable generation, assuming that ${\pmb\rho}$ is site-uncorrelated, positive (component by component) Gaussian,  thus represented by
\begin{eqnarray}
{\cal S}^{(White)}({\pmb \rho})\equiv\left\{\begin{array}{cc}
\sum_{i\in {\cal G}_u} (\rho_i/\overline{\rho}_i-1)^2, & \forall i\in{\cal G}_r:\quad \rho_i>0\\
+\infty, & \mbox{otherwise},\end{array}\right.
\label{Gauss}
\end{eqnarray}
where one chooses $\overline{\pmb \rho}=(\overline{\rho}_i|i\in{\cal G}_u$ in a way that the total of typical renewable generation, $\sum_{i\in {\cal G}_u}\overline{\rho}_i$, corresponds to $10\%$, $20\%$ and $30\%$ (for the three test cases respectively) of the standard generation of  the RTS-96 base case. Selecting $\overline{\pmb \rho}$ we also make sure that the resulting configurations are all in the regime where no transmission or generation constrains are violated, i.e. $\overline{\pmb \rho}\in{\cal D}_{int}$.  Note,  that the choice of the positive Gaussian and site-uncorrelated distribution is made here for illustrative purposes only.  Actual correlations of wind will be more elaborate and interesting. However, one expects that the realistic, ${\cal S}({\pmb \rho})$, representing actual wind forecast, will still be a convex function of ${\pmb \rho}$, thus making application of our algorithm to the more realistic situation as straightforward as for the synthetic positive Gaussian case discussed here.

The results of our numerical tests, which main goal was to illustrate utility of the algorithm as of an exact and fast tool, are shown in Figs.~(\ref{fig:3},\ref{fig:6},\ref{fig:9}) for the three levels of the renewable penetration respectively.  For each example we present two figures showing in, (a) configuration of renewable generation  ${\pmb\rho}$ for the top three leading instantons in comparison with the base case, and in (b) structure of the top three instantons (renewable sites with the largest values of $\rho/\overline{\rho}$ marked) and respective saturated edges. A complete instanton analysis of an instance (graph+distribution) is very fast, it takes few seconds on a laptop.

Here is the summary and brief discussion of the simulation results:
\begin{itemize}
\item In the cases correspondent to $10\%$ and $30\%$ of renewable penetration the top instantons are well localized on a site/node,  in a sense that production at this single site is significantly larger than the typical value while deviation in production at the other renewable sites are much weaker. On the contrary, the top instanton is of a de-localized type in the intermediate case of $20\%$ of renewable penetration. Note that the cost of the top instanton in the intermediate $20\%$ case , $S({\pmb \rho})=1.04$, is significantly lower than in the $10\%$ case, $S({\pmb \rho})=11.28$ and also smaller than in the $30\%$ case, $S({\pmb \rho})=2.49$. This non-monotonicity of the cost with increase of the renewable penetration illustrates the generally very important point that addition of renewables can stress (transition from the $10\%$ case to the ($30\%$ case) or de-stress (transition from the $10\%$ case to $20\%$ case) the network in terms of stochastic transmission congestion, depending on how it is done.

\item We also monitor the value of $\alpha$ at the instantons, and observe that it decreases with increase of the level of renewables: $0.75$, $0.63$ and $0.54$ for the top instantons in the $10\%$, $20\%$ and $30\%$ cases respectively. This expresses the fact that the addition of renewables translates into reduction of the standard generation.

\item There is always one saturated edge per instanton. \footnote{Note that this observation does not violate the $N-1$ contingency requirement (a security measure enforcing that the system has a feasible solution for any one of the $N$ edges of the graph removed) simply because the condition is enforced only at the equilibrium point.} Somehow remarkably, this saturated edge is always one of the ``old" edges (present in the base structure), and it is also positioned in majority of cases relatively far from the renewable sites which over-generate at the instanton. This observation emphasizes non-locality (and thus intrinsic difficulty) of the problem.

\item The standard generation constraints are not violated at the instantons,  i.e. for each of the instantons found, $\underline{\alpha}=0<\alpha<\overline{\alpha}=2$. $\alpha$ is unity at the base case and it (naturally) decreases with the level of renewable generation increase.

\end{itemize}

\section{Conclusions and Path Forward}
\label{sec:Summary}

Summarizing, the main results of this work are:
\begin{itemize}
\item We posed the problem of discovering extreme stochastic events in power grids, the instanton, associated with transmission overflows caused by fluctuations of renewable generation.

\item We showed that the aforementioned problem is computationally tractable within the DC power flow setting and with a low parametric linear control of standard generation, for example of the proportional type.

\item We illustrated algorithmic utility and efficiency of our newly suggested algorithm on example of the IEEE RTS-96 grid with $10\%$, $20\%$ and $30\%$ of added renewable generation. Main conclusions of our numerical tests are: (a) Addition of renewables may lead to significant increase of the probability of failures (destructive effect),  but it may also help to make the network to become prone to failures (constructive effect),  depending on quantitative details of how the grid extension is done. 
    (b) The instanton configurations represent global correlations in the graph, which shows itself in the fact that the (single) overloaded line is typically relatively far from the over-producing renewable sites. (c) Our algorithm is very efficient computationally in providing the exact assessment of the gain (or loss) associated with addition of renewables.
\end{itemize}

We plan to continue this work on discovering extreme stochastic events in the power grids efficiently.
Our main goal here is to design a reliable predictive tool, {\it extreme stochastic event toolbox},  capable to provide awareness (fast guidance to utility operator) in terms of predicting dangerous extreme stochastic events associated with the renewable generation. More specifically,  this work will be continued along the following lines.
\begin{itemize}
\item First of all, we will be testing scalability of the approach for larger,  continental scale, transmission networks. Our task here is to approach complexity which would scale linearly with the size of the system. (This may be achieved,  e.g. by replacing standard convex optimization solvers,  by their linear scaling and distributed proxies.)

\item Our modeling of wind statistics needs to be more realistic.  We plan to apply our algorithm to wind data taken from a realistic forecast, e.g. of the type available at \cite{NREL-data-sets}, in particular accounting for long spatial (usually,  at least hundreds of kilometers long) correlations between different sites. Our future test beds will include models of ERCOT (with actual or planned wind farms in Western Texas),  as well as models of other wind-rich parts of US power grids.

\item We will incorporate into the (so far static) scheme dynamic effects and actual (time series) measurements of the wind intensity. This will require extending the instanton approach to account for temporal Lagrangian correlations in the cost function, integrated in time over pre-history,  and enforcing the transmission and generation conditions not only instantaneously but also over the (dicretized) time horizon.

\item Our model of proportional control is convenient for analysis, but it matches the details of droop control and of regulation response only schematically.  We will modify the model to more fully represent these actions, in particular accounting for generation limits.

\item We envision incorporating the instanton analysis into control schemes,  in particular in the tertiary (balance) control, for example penalizing top instanton configurations (and their vicinities) in the modified optimum power flows. Another interesting control option is to fit the instanton framework into an adversarial process, whereby any given control action can be matched by a corresponding instanton.	This yields an iterative process (with each iteration roughly like the problem solved in the paper), which produces robust control actions and significant events, at the same time.

\item We will also work on extending the algorithm to the case of AC power flows (which is especially important in terms of accounting for additional issues related to voltage variations), e.g. utilizing new advances in related optimization techniques \cite{10LL}. We will consider  other schemes of transmission and voltage control,  in particular related to demand response \cite{11CH}, line switching \cite{09HOFO}, controlling DC tie-lines, capacitor banks,  phase shifters and related \cite{94Kun}. We also plan to account for other type of stochastic issues, e.g. related to dynamic stability \cite{89DH} and voltage collapse \cite{85VWC}.
\end{itemize}
Finally,  once the extreme stochastic event toolbox is developed, we envision using it (as a black box) for developing new planning and control schemes for smart grids of the future, for example in the spirit of the general approach of \cite{10BBT}. We envision using this toolbox to solve related problems of discovering interdiction attacks on power grids \cite{04SWB,08BV}, and analyzing, controlling and preventing cascading failures \cite{07DCLN,10PTC}.

\section{Acknowledgements}

We are thankful to D. Bienstock and P. Parrilo for very valuable and useful comments on theoretical and algorithmic aspects of the underlying optimization problem, to three anonymous referees for their remarks and comments which helped us to improve the presentation quality and discussion of limitations of our approach/technique, and to the participants of the ``Optimization and Control for Smart Grids" LDRD DR project at Los Alamos and Smart Grid Seminar Series at CNLS/LANL, and especially to S. Backhaus, R. Bent and K. Turitsyn, for multiple fruitful discussions.

\bibliographystyle{IEEETran} 
\bibliography{instanton_grid_new}

\begin{thebibliography}{10}
\providecommand{\url}[1]{#1}
\csname url@rmstyle\endcsname
\providecommand{\newblock}{\relax}
\providecommand{\bibinfo}[2]{#2}
\providecommand\BIBentrySTDinterwordspacing{\spaceskip=0pt\relax}
\providecommand\BIBentryALTinterwordstretchfactor{4}
\providecommand\BIBentryALTinterwordspacing{\spaceskip=\fontdimen2\font plus
\BIBentryALTinterwordstretchfactor\fontdimen3\font minus
  \fontdimen4\font\relax}
\providecommand\BIBforeignlanguage[2]{{%
\expandafter\ifx\csname l@#1\endcsname\relax
\typeout{** WARNING: IEEEtran.bst: No hyphenation pattern has been}%
\typeout{** loaded for the language `#1'. Using the pattern for}%
\typeout{** the default language instead.}%
\else
\language=\csname l@#1\endcsname
\fi
#2}}

\bibitem{09EDR}
\BIBentryALTinterwordspacing
``Description of the ``smart grid" at the website of the doe office of
  electricity delivery \& reliability.'' [Online]. Available:
  \url{http://www.oe.energy.gov/smartgrid.htm}
\BIBentrySTDinterwordspacing

\bibitem{09EERE}
\BIBentryALTinterwordspacing
``Description of plans, implementation, and results of the doe office of energe
  efficiency \& renewable energy.'' [Online]. Available:
  \url{http://www1.eere.energy.gov/pir/about.html}
\BIBentrySTDinterwordspacing

\bibitem{20wind}
DOE/GO-102008-2567, ``20 percent of wind energy by 2030 report: Increasing wind
  energy's contribution to us electricity supply,'' pp. 1--248, 2008.

\bibitem{64Lif}
I.~Lifshitz, ``The energy spectrum of disordered systems (energy spectrum of
  disordered solid with consideration of lattice vibration spectral density of
  crystal),'' \emph{Advances in Physics}, vol.~13, pp. 483--536, 1964.

\bibitem{66ZL}
J.~Zittartz and J.~S. Langer, ``Theory of bound states in a random potential,''
  \emph{Phys. Rev.}, vol. 148, no.~2, pp. 741--747, Aug 1966.

\bibitem{96FKLM}
G.~Falkovich, I.~Kolokolov, V.~Lebedev, and A.~Migdal, ``Instantons and
  intermittency,'' \emph{Phys. Rev. E}, vol.~54, no.~5, pp. 4896--4907, Nov
  1996.

\bibitem{97Che}
M.~Chertkov, ``Instanton for random advection,'' \emph{Phys. Rev. E}, vol.~55,
  no.~3, pp. 2722--2735, Mar 1997.

\bibitem{08CS}
M.~Chertkov and M.~Stepanov, ``An efficient pseudocodeword search algorithm for
  linear programming decoding of {L}{D}{P}{C} codes,'' \emph{Information
  Theory, IEEE Transactions on}, vol.~54, no.~4, pp. 1514--1520, April 2008.

\bibitem{11CPS}
M.~Chertkov, F.~Pan, and M.~Stepanov, ``Predicting failures in power grids: The
  case of static overloads,'' \emph{IEEE Transactions on Smart Grids}, vol.~99,
  p.~1, 2011.

\bibitem{94Kun}
P.~Kundur, \emph{Power System Stability and Control}.\hskip 1em plus 0.5em
  minus 0.4em\relax New York, NY, USA: McGraw-Hill, 1994.

\bibitem{96WW}
A.~J. Wood and B.~F. Wollenberg, \emph{Power Generation Operation And Control},
  2nd~ed., New York, 1996.

\bibitem{Weibull}
\BIBentryALTinterwordspacing
``Describing wind variations: Weibull distribution.'' [Online]. Available:
  \url{http://wiki.windpower.org/index.php/The_Weibull_distribution}
\BIBentrySTDinterwordspacing

\bibitem{04CCSV}
V.~Chernyak, M.~Chertkov, M.~Stepanov, and B.~Vasic, ``Instanton method of
  post-error-correction analytical evaluation,'' \emph{Information Theory
  Workshop, 2004. IEEE}, pp. 220--224, Oct. 2004.

\bibitem{05SCCV}
M.~G. Stepanov, V.~Chernyak, M.~Chertkov, and B.~Vasic, ``Diagnosis of
  weaknesses in modern error correction codes: A physics approach,''
  \emph{Phys. Rev. Lett.}, vol.~95, no.~22, p. 228701, Nov 2005.

\bibitem{AMOEBA}
J.~Nelder and R.~Mead, ``A simplex method for function minimization,''
  \emph{Computer Journal}, vol.~7, no.~4, pp. 308--313, 1965.

\bibitem{NumRec}
W.~Press, B.~Flannery, S.~Teukolsky, and W.~Vetterling, \emph{Numerical
  Recipes: The Art of Scientific Computing}.\hskip 1em plus 0.5em minus
  0.4em\relax New York: Cambridge University Press, 1986.

\bibitem{97Bri}
\BIBentryALTinterwordspacing
W.~Briec, ``Minimum distance to the complement of a convex set: Duality
  result,'' \emph{Journal of Optimization Theory and Applications}, vol.~93,
  pp. 301--319, 1997, 10.1023/A:1022697822407. [Online]. Available:
  \url{http://dx.doi.org/10.1023/A:1022697822407}
\BIBentrySTDinterwordspacing

\bibitem{96FP}
K.~Fukuda and A.~Prodon, ``Double description method revisited,'' in
  \emph{Combinatorics and Computer Science, volume 1120 of Lecture Notes in
  Computer Science}, M.~Deza, R.~Euler, and I.~Manoussakis, Eds.\hskip 1em plus
  0.5em minus 0.4em\relax Springer-Verlag, 1996, p. 91–111.

\bibitem{00Avi}
D.~Avis, ``A revised implementation of the reverse search vertex enumeration
  algorithm,'' in \emph{Polytopes — Combinatorics and Computation}, K.~G. and
  G.~M. Ziegler, Eds.\hskip 1em plus 0.5em minus 0.4em\relax Birkhauser-Verlag,
  2000, p. 177–198.

\bibitem{96-RTS}
C.~Grigg, P.~Wong, P.~Albrecht, R.~Allan, M.~Bhavaraju, R.~Billinton, Q.~Chen,
  C.~Fong, S.~Haddad, S.~Kuruganty, W.~Li, R.~Mukerji, D.~Patton, N.~Rau,
  D.~Reppen, A.~Schneider, M.~Shahidehpour, and C.~Singh, ``The ieee
  reliability test system-1996. a report prepared by the reliability test
  system task force of the application of probability methods subcommittee,''
  \emph{Power Systems, IEEE Transactions on}, vol.~14, no.~3, pp. 1010 --1020,
  aug 1999.

\bibitem{NREL-data-sets}
\BIBentryALTinterwordspacing
``Nrel wind integration data sets.'' [Online]. Available:
  \url{http://www.nrel.gov/wind/integrationdatasets/}
\BIBentrySTDinterwordspacing

\bibitem{10LL}
\BIBentryALTinterwordspacing
J.~Lavaei and S.~H. Low, ``Zero duality gap in optimal power flow problem,''
  2010. [Online]. Available:
  \url{https://www.cds.caltech.edu/~lavaei/ACC2011_1.pdf}
\BIBentrySTDinterwordspacing

\bibitem{11CH}
D.~Callaway and I.~Hiskens, ``Achieving controllability of electric loads,''
  \emph{Proceedings of the IEEE}, vol.~99, no.~1, pp. 184 --199, 2011.

\bibitem{09HOFO}
K.~W. Hedman, R.~P. O'Neill, E.~B. Fisher, and S.~S. Oren, ``Optimal
  transmission switching with contingency analysis,'' \emph{IEEE Transactions
  Power Systems}, vol.~24, no.~3, 2009.

\bibitem{89DH}
I.~Dobson and H.-D. Chiang, ``Towards a theory of voltage collapse in electric
  power systems,'' \emph{Syst. Control Lett.}, vol.~13, no.~3, pp. 253--262,
  1989.

\bibitem{85VWC}
P.~Varaiya, F.~Wu, and R.-L. Chen, ``Direct methods for transient stability
  analysis of power systems: Recent results,'' \emph{Proceedings of the IEEE},
  vol.~73, no.~12, pp. 1703 -- 1715, 1985.

\bibitem{10BBT}
R.~Bent, A.~Berscheid, , and G.~Toole, ``Transmission network expansion
  planning with simulation optimization,'' in \emph{Proceedings of the
  Twenty-Fourth AAAI Conference on Artificial Intelligence (AAAI 2010)}, 2010.

\bibitem{04SWB}
J.~Salmeron, K.~Wood, and R.~Baldick, ``Analysis of electric grid security
  under terrorist threat,'' \emph{Power Systems, IEEE Transactions on},
  vol.~19, no.~2, pp. 905--912, May 2004.

\bibitem{08BV}
\BIBentryALTinterwordspacing
D.~Bienstock and A.~Verma, ``The n-k problem in power grids: New models,
  formulations and computations,'' \emph{to appear in SIAM J. of Optimization},
  2010. [Online]. Available: \url{http://www.columbia.edu/~dano/papers/nmk.pdf}
\BIBentrySTDinterwordspacing

\bibitem{07DCLN}
I.~{Dobson}, B.~A. {Carreras}, V.~E. {Lynch}, and D.~E. {Newman}, ``{Complex
  systems analysis of series of blackouts: Cascading failure, critical points,
  and self-organization},'' \emph{Chaos}, vol.~17, no.~2, pp. 026\,103--+, June
  2007.

\bibitem{10PTC}
\BIBentryALTinterwordspacing
R.~{Pfitzner}, K.~{Turitsyn}, and M.~{Chertkov}, ``{Statistical Classification
  of Cascading Failures in Power Grids},'' 2010. [Online]. Available:
  \url{http://arxiv.org/abs/1012.0815}
\BIBentrySTDinterwordspacing

\end{thebibliography}

\end{document}